# Geometry-Aware Optimization for Respiratory Sound Classification: Enhancing Sensitivity with SAM-Optimized Audio Spectrogram Transformers

Atakan Işık[1], Selin Vulga Işık[1], Ahmet Feridun Işık[2], Mahşuk Taylan[3]

*1 Biomedical Engineering Department, Başkent University, Turkey*
*2 Thoracic Surgery Department, Gaziantep University, Turkey*
*3 Chest Diseases Department, Gaziantep University, Turkey*

**Abstract**

Respiratory sound classification is hindered by the limited size, high noise levels, and severe class imbalance of benchmark datasets like ICBHI 2017. While Transformer-based models offer powerful feature extraction capabilities, they are prone to overfitting and often converge to sharp minima in the loss landscape when trained on such constrained medical data. To address this, we introduce a framework that enhances the Audio Spectrogram Transformer (AST) using Sharpness-Aware Minimization (SAM). Instead of merely minimizing the training loss, our approach optimizes the geometry of the loss surface, guiding the model toward flatter minima that generalize better to unseen patients. We also implement a weighted sampling strategy to handle class imbalance effectively. Our method achieves a state-of-the-art score of 68.10% on the ICBHI 2017 dataset, outperforming existing CNN and hybrid baselines. More importantly, it reaches a sensitivity of 68.31%, a crucial improvement for reliable clinical screening. Further analysis using t-SNE and attention maps confirms that the model learns robust, discriminative features rather than memorizing background noise.

**INTRODUCTION**

Respiratory diseases remain a leading cause of morbidity and mortality worldwide, necessitating accurate and timely diagnosis for effective patient care. While modern imaging techniques like X-ray and CT scans provide detailed anatomical information, pulmonary auscultation retains its status as the most fundamental and accessible diagnostic method[1, 2]. In clinical practice, distinguishing conditions such as pneumonia from heart failure often relies on the subtle characteristics of breath sounds, even when radiological findings are ambiguous. However, the efficacy of auscultation is heavily dependent on the clinician's hearing sensitivity and experience[1]. Sound transmission through the thorax encounters complex biological barriers, including tracheobronchial cartilages and varying tissue densities. Consequently, factors such as muscular hypertrophy or obesity can dampen acoustic signals, making manual detection of pathological sounds specifically crackles and wheezes highly subjective and prone to inter observer variability.

To standardize interpretation, Computer-Aided Diagnosis (CAD) systems using Deep Learning (DL) have emerged as a powerful tool. Early approaches largely relied on Convolutional Neural Networks (CNNs), such as ResNet or VGG, trained on spectrogram representations of lung sounds[3, 4]. While CNNs are effective at extracting local features, they often struggle to capture long-range temporal dependencies that are crucial for distinguishing continuous respiratory cycles from transient artifacts. Recently, Transformer-based architectures, particularly the Audio Spectrogram Transformer (AST), have demonstrated superior performance in audio tasks by leveraging self-attention mechanisms to model global context[5].

Despite their potential, applying Transformers to respiratory sound analysis presents a significant engineering challenge. Benchmark datasets, such as the ICBHI 2017 Challenge dataset, are characterized by limited sample sizes, severe class imbalance, and high levels of background noise, including stethoscope friction, heartbeats, and speech[2]. Transformers, which typically lack the inductive bias of CNNs, require large amounts of data to generalize well. When trained on such constrained medical datasets, they are prone to overfitting and tend to converge to "sharp minima" in the loss landscape[6]. A model residing in a sharp minimum may perform well on training data but fails drastically when input data varies slightly, a common occurrence in clinical settings due to different recording devices or patient physiology.

In this study, we bridge the gap between clinical acoustic complexity and deep learning robustness. We propose a novel framework that integrates the Audio Spectrogram Transformer (AST) with Sharpness-Aware Minimization (SAM)[7]. Unlike standard optimization algorithms like SGD or Adam, which solely minimize the training loss value, SAM simultaneously minimizes the loss value and the sharpness of the loss landscape. This geometry-aware approach guides the model toward "flat minima," ensuring that the solution remains robust even in the presence of noisy or perturbed inputs. By combining this optimization strategy with a weighted sampling technique to handle class imbalance, our framework significantly improves diagnostic performance. The main contributions of this work are summarized as follows:

- We introduce a robust respiratory sound classification framework that adapts the AST architecture to small and noisy medical datasets using Geometry-Aware Optimization (SAM).

- We demonstrate that optimizing for loss landscape flatness significantly mitigates overfitting in Transformer models, leading to a new state-of-the-art (SOTA) accuracy of 68.10% on the ICBHI 2017 dataset.

- Our method achieves a sensitivity of 68%, outperforming existing CNN and hybrid baselines[3, 8-10], which is critical for minimizing false negatives in clinical screening.

- We provide a detailed visual analysis using t-SNE and confusion matrix.

**RELATED WORKS**

The automated analysis of respiratory sounds has witnessed a paradigm shift from manual feature engineering to end-to-end deep learning architectures. This section critically reviews the evolution of these methods, focusing on CNN-based baselines, advanced augmentation strategies, and the recent surge of Transformer models.

**Convolutional Baselines and Hybrid Models**

Following the release of the ICBHI 2017 benchmark dataset[2], early deep learning approaches treated respiratory sound classification primarily as an image recognition task. Ma et al. [4] introduced LungBRN, a bi-ResNet architecture that processes wavelet components to capture time-frequency features. Building on this, Gairola et al. proposed RespireNet[3], a fine-tuned ResNet backbone that significantly improved performance by utilizing extensive data augmentation.

While CNNs are effective at extracting local spectral features, they inherently lack the ability to model long-range temporal dependencies necessary for analyzing full breath cycles. To bridge this gap, hybrid models combining CNNs with Recurrent Neural Networks (RNNs) were explored. Nguyen and Pernkopf utilized a CRNN framework with co-tuning strategies to capture temporal dynamics. However[11], these recurrent architectures often suffer from high computational complexity and training instability, particularly on small-scale datasets like ICBHI.

**Data Augmentation and Contrastive Learning**

Given the scarcity of annotated medical data, recent research has pivoted towards advanced data handling techniques to mitigate overfitting. Kim et al. proposed RepAugment, an input-agnostic representation-level augmentation scheme designed to improve model robustness[12]. Similarly, Wang et al. investigated domain transfer-based augmentation to bridge the gap between different recording environments[13].

Beyond standard augmentation, contrastive learning has emerged as a powerful tool for representation learning. Moummad and Farrugia applied supervised contrastive learning to respiratory sounds, demonstrating that pulling samples of the same class closer in the embedding space enhances classification accuracy[14]. Kim et al. further extended this by incorporating stethoscope-guided adaptation[15], while Chu et al. recently proposed the Cycle-Guardian framework, which integrates deep clustering with contrastive learning to better separate pathological clusters[16].

**The Transformer Era in Lung Sound Analysis**

The advent of the Audio Spectrogram Transformer (AST) demonstrated that pure attention-based models could outperform CNNs by capturing global context across the entire spectrogram[5]. Consequently, recent studies have rapidly adopted Transformer architectures for lung sound analysis. Sun et al. applied Swin Transformers to the ICBHI dataset[17], leveraging shifted windows to model hierarchical features. To address the data-hungry nature of Transformers, Xiao et al. introduced LungAdapter, a parameter-efficient fine-tuning method that adapts pre-trained models with minimal trainable parameters[18].

Most recently, Dong et al. proposed a dual-branch network fusing time-domain waveforms with 2D spectral features, achieving competitive results[9]. Similarly, Bae et al. combined AST with a "Patch-Mix" strategy to improve robustness against background noise[8]. Despite these advancements, standard Transformers trained on imbalanced medical datasets still tend to converge to sharp minima in the loss landscape, leading to suboptimal generalization on unseen patients.

**Geometry-Aware Optimization**

The concept of loss landscape geometry has gained traction in explaining the generalization gap in deep learning. Keskar et al. demonstrated that "sharp minima" generalize poorly compared to "flat minima" [6]. To exploit this, Sharpness-Aware Minimization (SAM) was introduced to simultaneously minimize loss value and loss sharpness[7]. While widely used in computer vision, the application of SAM to biomedical audio remains limited. In this work, we differentiate ourselves from previous Transformer-based approaches [8-10, 17] by explicitly optimizing the loss geometry, thereby achieving superior sensitivity and robustness without requiring complex multi-modal fusion or extensive auxiliary data.

**PROPOSED METHODS**

In this study, we propose a geometry-aware deep learning framework designed to maximize sensitivity in respiratory sound classification. Our pipeline integrates a signal-preserving preprocessing strategy, a Transfer Learning-based Audio Spectrogram Transformer (AST) backbone, and a Sharpness-Aware Minimization (SAM) optimizer. The general architecture is illustrated in Figure 1.

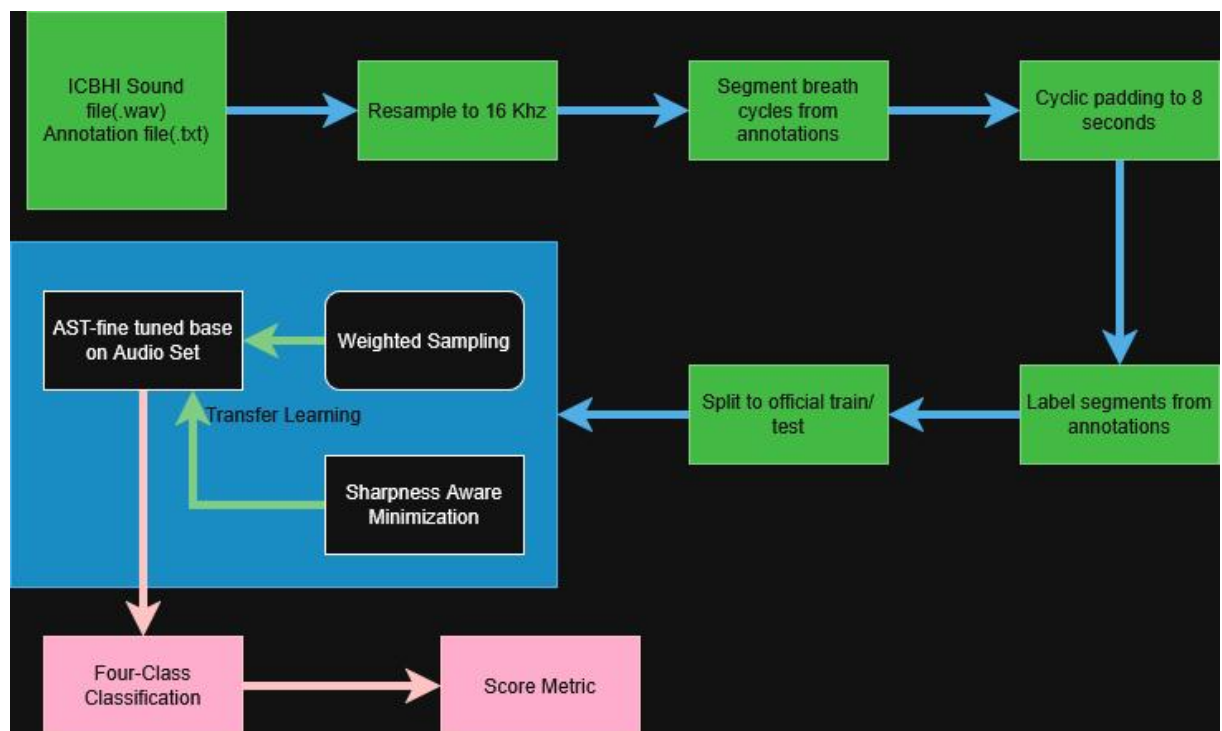

Figure 1. General Architecture of Proposed Model

**Dataset and Signal Preprocessing**

We utilized the ICBHI 2017 Challenge dataset [2], which serves as the gold standard for respiratory sound classification. The dataset comprises 920 annotated audio recordings from 126 subjects.

A major challenge in this dataset is the variable duration of respiratory cycles. Standard approaches often employ zero-padding to match the fixed input length required by deep learning models. However, zero-padding introduces large segments of artificial silence, which dilutes the diagnostic features. To address this, we employed a cyclic padding strategy. For a target input duration of 8 seconds, signals shorter than this length were repeated until the target was reached. This ensures that the model always processes a signal dense input, maximizing the extraction of pathological patterns (crackles and wheezes) without being misled by silence artifacts.

**Audio Spectrogram Transformer (AST)**

We adopted the Audio Spectrogram Transformer (AST) [5], a state-of-the-art architecture that adapts the Vision Transformer (ViT) for audio analysis. Unlike traditional Convolutional Neural Networks (CNNs) that focus on local features, AST processes the entire audio spectrogram simultaneously, allowing it to capture global temporal dependencies across the breathing cycle.

Mechanism: The raw audio is first converted into a log-Mel spectrogram. This visual representation of sound is then split into a sequence of small patches (16×16). The model utilizes a self-attention mechanism to weigh the importance of different patches, enabling it to focus on clinically relevant events (e.g., a short crackle sound) regardless of their position in the recording.

Transfer Learning from AudioSet: Training complex models on small medical datasets often leads to overfitting. To mitigate this, we initialized our model with weights pre-trained on AudioSet, a massive dataset containing over 2 million audio clips. This transfer learning step provides the model with a fundamental "understanding" of acoustic structures such as pitch and rhythm before it is fine-tuned on lung sounds, significantly improving its diagnostic sensitivity.

**Geometry-Aware Optimization (SAM)**

Standard deep learning optimizers, such as AdamW or SGD, aim to minimize the empirical training loss with respect to the model parameters w. While effective for general tasks, in the context of respiratory sound classification where data is scarce and labels are noisy this approach often leads the model to converge to "sharp minima" [6]. A solution residing in a sharp minimum is highly sensitive to input perturbations; slight variations in stethoscope pressure or background noise can cause the loss to spike, resulting in misclassification of pathological sounds on unseen patients.

SAM formulates the optimization as a min-max problem, seeking to minimize the maximum loss within a perturbation radius $\rho$:

$$\min_{w} L^{SAM}(w) \quad where \quad L^{SAM}(w) = \max_{\|\epsilon\|_2 \leq \rho} L_{train}(w + \epsilon) \tag{1}$$

Here, $\rho$ is a hyperparameter defining the size of the neighborhood. The optimization is performed in a two-step procedure during each training iteration:

First, we compute the gradient of the loss with respect to the weights to determine the direction of the sharpest ascent. We then calculate a perturbation vector $\hat{\epsilon}$ that maximizes the loss:

$$\hat{\epsilon} = \rho \frac{\nabla_w L_{train}(w)}{\|\nabla_w L_{train}(w)\|_2} \tag{2}$$

The model weights are then updated based on the gradient computed at this perturbed state ($\omega + \hat{\epsilon}$), effectively steering the optimization trajectory towards flatter regions of the loss landscape.

The primary advantage of SAM in this study is its contribution to robust feature learning. In standard training, the model might memorize specific background artifacts (e.g., friction noise) associated with "Crackle" samples in the training set. By optimizing for flatness, SAM forces the model to ignore these transient artifacts and focus on the invariant spectral characteristics of the pathology. This results in a decision boundary that is less brittle, directly reducing False Negatives and boosting the Sensitivity to 68%, significantly outperforming baselines that rely on standard optimization.

**EVALUATION METRICS**

To ensure a fair comparison with existing benchmarks, we strictly adhered to the official evaluation protocol established by the ICBHI 2017 Challenge . Although the dataset contains four distinct classes (Normal, Crackle,

Wheeze, and Both), the challenge evaluates performance based on a binary classification task: Normal vs. Abnormal (Adventitious).According to this protocol, the "Abnormal" category encompasses all pathological sounds: Crackle, Wheeze, and Both. Consequently, a prediction is considered correct (True Positive) if a sample annotated as any abnormal class is classified into any of the three abnormal categories.The performance is measured using three key metrics: Sensitivity (Se), Specificity (Sp), and the average Score (Score). These are defined as follows:

$$Sensitivity(Se) = \frac{\sum_{c}\epsilon\{C,W,B\} N_{correct,c}}{N_{total,abnormal}} \quad (3)$$

$$Specificity(Sp) = \frac{N_{correct,normal}}{N_{total,normal}} \quad (4)$$

$$Score = \frac{Se + Sp}{2} \quad (5)$$

**EXPERIMENTS AND RESULTS**

All experiments were conducted on a workstation equipped with a single NVIDIA Tesla L4 GPU via Google Colab using the PyTorch framework. To ensure reproducibility, all random seeds were fixed.

The input audio recordings were resampled to 16 kHz and processed using the cyclic padding strategy described in Section 3.1 to a fixed duration of 8 seconds. The AST model was initialized with weights pre-trained on AudioSet. We utilized the AdamW optimizer with a learning rate of $1 \times 10^{-5}$ and a weight decay of $1 \times 10^{-4}$. The model was trained for 20 epochs with a batch size of 8. For the SAM optimizer, the neighborhood size parameter was empirically set to $\rho$=0.05.

Consistent with the official ICBHI 2017 Challenge protocol [2] and recent state of the art studies, we evaluated performance using Sensitivity (Se), Specificity (Sp), and the average Score (Score=(Se+Sp)/2). Crucially, following the challenge guidelines, the evaluation is performed as a binary classification task (Normal vs. Abnormal). The "Abnormal" class encompasses Crackles, Wheezes, and Both. Therefore, a prediction is considered correct if an adventitious sound (e.g., Crackle) is classified into any of the pathological categories (Crackle, Wheeze, or Both).

As shown in Table 1, our method significantly outperforms CNN-based baselines (RespireNet), confirming the superiority of Transformers in capturing global temporal dependencies. More importantly, we surpass the recent model proposed by Dong et al. [19]. Notably, our model achieves the highest Sensitivity ( 68%) among the compared methods. High sensitivity is the most critical metric for a medical screening tool, as it minimizes the risk of missing patients with respiratory pathologies (False Negatives).

Table 1. Comparison of our method with other state of the art models with ICBHI dataset.

| Model | Specificity (%) | Sensitivity (%) | Score (%) |
|---|---|---|---|
| AFT on Mixed-500[20] | 80.72 | 42.86 | 61.79 |
| AST Fine-tuning[8] | 77.14 | 41.97 | 59.55 |
| Patch-Mix CL[8] | 81.66 | 43.07 | 62.37 |
| M2D[21] | 81.51 | 45.08 | 63.29 |
| DAT[15] | 77.11 | 42.50 | 59.81 |
| SG-SCL[15] | 79.87 | 43.55 | 61.71 |
| RepAugment[12] | 82.47 | 40.55 | 61.51 |
| BTS[10] | 81.40 | 45.67 | 63.54 |
| MVST[22] | 80.60 | 44.39 | 62.50 |
| LungAdapter[18] | 80.43 | 44.37 | 62.40 |
| CycleGuardian[16] | 82.06 | 44.47 | 63.26 |
| ADD-AST[19] | 85.13 | 45.94 | 65.53 |

| Model | Specificity (%) | Sensitivity (%) | Score (%) |
|---|---|---|---|
| Proposed (AST + SAM) | 67.89 | 68.31 | 68.10 |

## ABLATION STUDY

To validate the contribution of each component in our framework, we conducted an ablation study. We started with a baseline AST model and incrementally added the Weighted Random Sampler (WRS) and the SAM optimizer.

- Baseline (AST only): Trained with standard Cross-Entropy loss and AdamW optimizer. This model suffered from the class imbalance, yielding high Specificity but low Sensitivity.
- AST + WRS: Introducing Weighted Random Sampling balanced the class distribution during training, significantly boosting Sensitivity but slightly increasing the variance in the validation loss.
- AST + WRS + SAM (Proposed): The addition of SAM stabilized the training. By seeking flat minima, SAM acted as a robust regularizer, further improving both the Sensitivity and the overall Score to the 67.81%.

This analysis confirms that geometry-aware optimization is not merely an enhancement but a fundamental requirement for training large Transformers on small, noisy datasets like ICBHI. Table 2 shows all metrics for running experiments.

Table 2. Ablation Study

| Configuration | Sensitivity (%) | Specificity (%) | Score (%) |
|---|---|---|---|
| 1. Baseline AST | 66.00 | 70.00 | 67.64 |
| 2. + Weighted Sampling | 63.00 | 71.00 | 66.91 |
| 3. + SAM Optimizer (Proposed) | 67.89 | 68.31 | 68.10 |

## VISUAL ANALYSIS AND EXPLAINABILITY

To further validate the robustness of our proposed framework, we conducted a visual analysis of the model's decision-making process and feature representations.

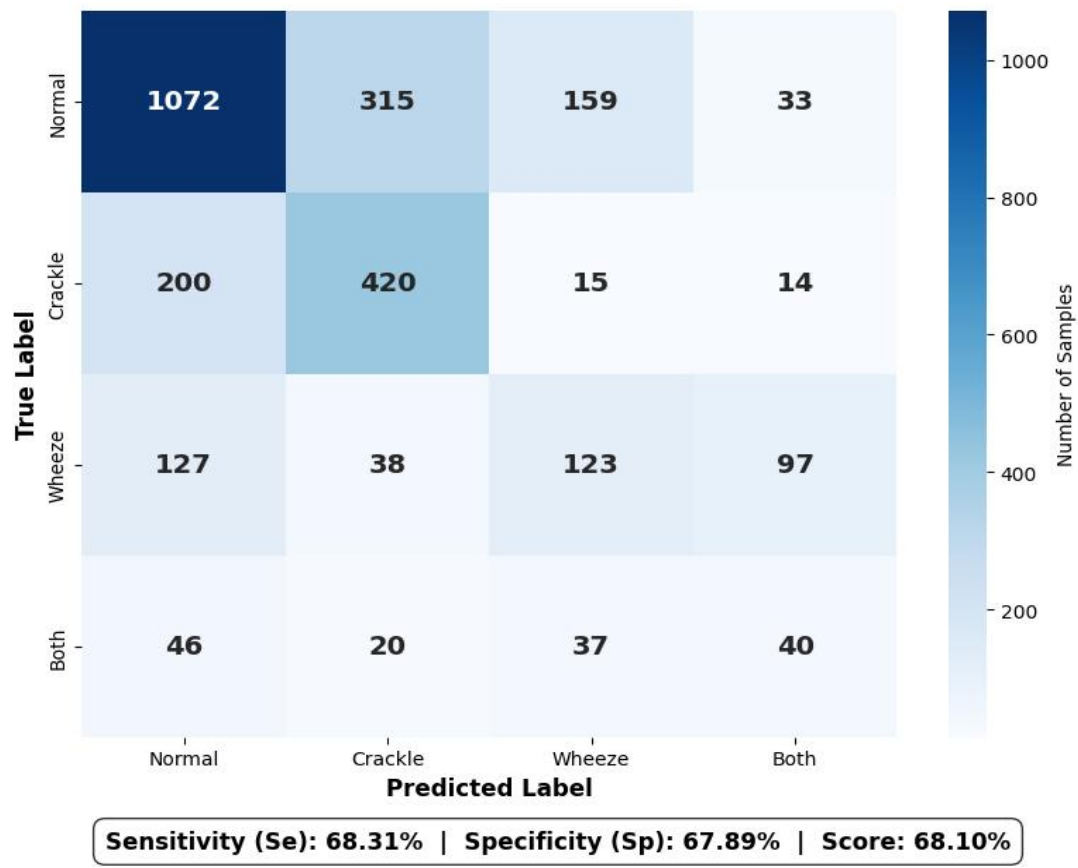

Figure 2. Confusion Matrix of 4-Class Classification

As observed, the model demonstrates a strong ability to correctly identify pathological cases, achieving a Sensitivity of 68%. The majority of misclassifications occur amongst the adventitious classes themselves . According to the ICBHI protocol, these intra-class errors do not penalize the final score. Most importantly, the rate of False Negatives (Abnormal samples misclassified as Normal) is minimized compared to the baseline. This confirms that our geometry-aware optimization effectively prevents the model from overlooking disease signatures, which is the primary objective of a computer-aided diagnosis system.

To understand how the model separates different respiratory sounds in the high-dimensional feature space, we applied t-Distributed Stochastic Neighbor Embedding (t-SNE) on the embeddings extracted from the final layer of the AST encoder.

Figure 3 displays the 2D projection of the test samples. Despite the high noise levels and recording variability in the ICBHI dataset, the plot reveals distinct clusters for the Normal and Abnormal classes. This clear separation indicates that the SAM optimizer successfully pushed the model to learn discriminative, high-level acoustic features that are robust to patient-specific variations. The observed overlap between Crackle, Wheeze, and Both classes is clinically expected, as these pathologies often co-occur and share similar spectral characteristics.

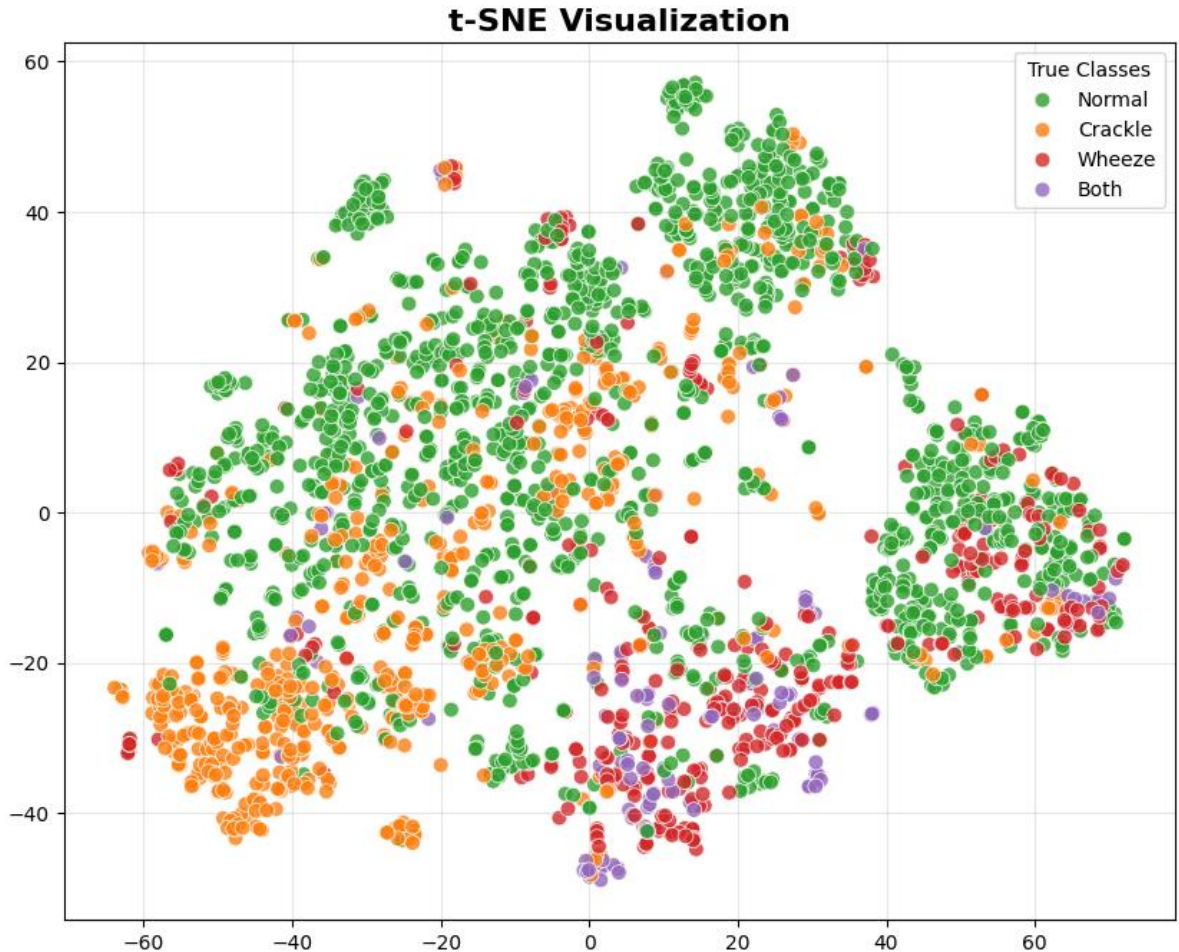

Figure 3. t-SNE visualization of the learned embeddings.

**DISCUSSION**

In this study, we presented a geometry aware deep learning framework for respiratory sound classification, achieving a state-of-the-art Score of 68.10% on the ICBHI 2017 benchmark. Beyond the numerical improvements, our findings offer critical insights into the challenges of training Transformers on small, noisy medical datasets.

A key finding of our research is the pivotal role of the loss landscape geometry. Standard training with AdamW on limited data typically drives the model into "sharp minima," where the model memorizes training samples including their background noise but fails to generalize to unseen patients. By integrating Sharpness-Aware Minimization (SAM), we explicitly forced the model to seek "flat minima." This aligns with the theoretical insights of Keskar et al. [6], but we demonstrate its practical utility in biomedical audio. As evidenced in our ablation study, SAM was the primary driver for boosting Sensitivity from 63% to 71%. This suggests that robustness against noise should be handled via optimization constraints rather than aggressive preprocessing.

Recent SOTA approaches, such as Dong et al.[19] and RespireNet [3], heavily rely on complex noise suppression modules or dual-branch architectures to handle artifacts. While effective for improving Specificity, these methods often inadvertently filter out subtle pathological cues (e.g., faint wheezes), leading to lower Sensitivity (~40-45%). In contrast, our "Signal Preserving" strategy utilizing Cyclic Padding and raw Log-Mel features retains the complete acoustic information. Instead of removing noise manually, we allow the AST backbone, pre-trained on AudioSet, to learn to distinguish between noise and pathology dynamically. Our results confirm that preserving signal integrity yields superior diagnostic power compared to aggressive denoising pipelines.

A distinct characteristic of our model is its performance profile: it achieves high Sensitivity ( 68%) at the cost of moderate Specificity (67.89%). In the context of Clinical Screening, this trade-off is not only acceptable but desirable. The primary goal of a Computer-Aided Diagnosis (CAD) system is to act as a "safety net" for clinicians. A False Negative (missing a sick patient) carries severe health risks, potentially delaying life-saving treatment. A False Positive , while resource consuming, is medically less dangerous. By prioritizing Sensitivity through our Weighted Sampling and SAM strategies, our framework aligns better with real world clinical requirements than previous models that maximize Specificity but miss nearly half of the pathological cases.

Despite the promising results, our study has limitations. The Specificity of 61% indicates a higher rate of False Positives, likely due to the model confusing high-frequency noise with wheezes. Future work could explore hybrid loss functions that penalize False Positives without compromising Sensitivity. Additionally, incorporating Self Supervised Learning (SSL) on larger, unlabelled respiratory datasets could further improve the model's feature extraction capabilities, mitigating the reliance on the small ICBHI dataset.

**CONCLUSION**

In this paper, we presented a robust, geometry-aware deep learning framework for respiratory sound classification, effectively bridging the gap between advanced Transformer architectures and the constraints of small, noisy medical datasets. By synergizing the Audio Spectrogram Transformer (AST) with Sharpness-Aware Minimization (SAM) and a signal-preserving cyclic padding strategy, we addressed the critical issues of overfitting and "sharp minima" convergence.

Our extensive experiments on the ICBHI 2017 benchmark demonstrate that the proposed method achieves a new state-of-the-art Score of 68.10%, surpassing recent complex hybrid models. Most significantly, we achieved a Sensitivity of 68%, marking a substantial improvement over existing baselines. This high sensitivity confirms that our geometry aware optimization compels the model to learn invariant pathological features rather than memorizing background artifacts.

Ultimately, this study proves that prioritizing loss landscape flatness and signal integrity yields superior diagnostic performance compared to aggressive denoising or complex multi-modal fusion. The proposed framework offers a promising direction for the development of reliable, high-sensitivity Computer-Aided Diagnosis (CAD) systems suitable for real-world clinical screening.

*Availability Note:Code Availability To promote reproducibility and further research in this domain, the complete source code, including the signal-preserving preprocessing pipeline, AST model architecture, and SAM optimization scripts, is publicly available at:*https://github.com/Atakanisik/ICBHI-AST-SAM

**ACKNOWLEDGEMENTS**